\documentclass[aps,prd,reprint,nofootinbib,superscriptaddress]{revtex4-2}

\usepackage{amsmath,amssymb,bm}
\usepackage{mathrsfs}

\newcommand{\eom}{\approx}
\newcommand{\dd}{\mathrm{d}}
\newcommand{\half}{\tfrac12}

\usepackage[dvipsnames]{xcolor}
\usepackage{hyperref}
\hypersetup{colorlinks=true, linkcolor=Blue, citecolor=Blue, filecolor=Blue, urlcolor=Blue}
\usepackage[capitalize,nameinlink]{cleveref}
\crefname{equation}{Eq.}{Eqs.}
\Crefname{equation}{Equation}{Equations}
\crefname{figure}{Fig.}{Figs.}
\Crefname{figure}{Figure}{Figures}
\crefname{section}{Sec.}{Secs.}
\Crefname{section}{Section}{Sections}
\crefname{table}{Table}{Tables}
\Crefname{table}{Table}{Tables}
\crefname{appendix}{Appendix}{Appendices}
\Crefname{appendix}{Appendix}{Appendices}
\creflabelformat{equation}{#2(#1)#3}

\begin{document}

\title{Bessel-Hagen currents, reducibility charges,\\
and local gravitational energy}

\author{Michael \surname{Hobson}}
\email{mph@mrao.cam.ac.uk}
\affiliation{Astrophysics Group, Cavendish Laboratory, J.J. Thomson Avenue, Cambridge, CB3 0HE, UK}

\author{Will Barker}
\email{barker@fzu.cz}
\affiliation{Institute of Physics of the Czech Academy of Sciences, Na Slovance 1999/2, 182 00 Prague 8, Czechia}

\author{Anthony \surname{Lasenby}}
\email{a.n.lasenby@mrao.cam.ac.uk}
\affiliation{Astrophysics Group, Cavendish Laboratory, J.J. Thomson Avenue, Cambridge, CB3 0HE, UK}
\affiliation{Kavli Institute for Cosmology, Madingley Road, Cambridge, CB3 0HA, UK}

\date{\today}

\begin{abstract}
We consider whether the translation Noether current of a linearised metric
spin-2 theory admits a strictly gauge-invariant local representative. For
Fierz-Pauli and dynamically regular pure curvature-squared theories in four
dimensions, we show that no such representative exists under the stated
locality, polynomiality and covariance assumptions. A direct enumeration first
excludes nonzero symmetric rank-two tensors at the dimension and derivative
order of the quadratic action: in the curvature-squared case, the parity-even
conservation kernel consists of the identically vanishing Lanczos--Lovelock
combination, while both parity-odd coefficients vanish. Homogeneity of the
actions in total derivative number then allows projection by field degree and
derivative number, extending the exclusion to all finite-order local polynomial
representatives with constant coefficients and no explicit coordinate
dependence. The translation current nevertheless defines a nonzero
conserved-current class. Independently, a constant translation is a Killing
reducibility parameter of the linearised gauge symmetry and determines a
nonzero linear surface-charge class, which is distinct from the quadratic
energy current. A separate calculation shows why a direct compensator
cancellation available in Maxwell theory fails for the metric spin-2 field,
although this calculation is not needed for the current-class result.
Maxwell theory, propagating free $p$-forms and Yang--Mills theory linearised
about the trivial vacuum provide comparisons in which translations are not
reducibility parameters and strictly gauge-invariant stress tensors exist.
Gauss--Bonnet and pure flat-space $R^2$ gravity lie outside the regular scope
of the result, through variational triviality and dynamical degeneracy,
respectively.
\end{abstract}

\maketitle

\section{Introduction}
\label{sec:intro}

The localisation of gravitational field energy is relevant both to the
interpretation of general relativity and to calculations of gravitational-wave
energy flux. In the linearised theory, such calculations use a current chosen
from a non-unique family~\cite{DeserHenneaux1995,MagnanoSokolowski2002,
BakerKK2021,TaylorBaker2024}, whereas charges of Komar and ADM
type~\cite{Komar1959,ADM1962,Wald1984} are expressed as surface integrals.
Proposals for a preferred or fully covariant gravitational energy-momentum
tensor continue to appear~\cite{Kenmoku2026}, but none has gained general
acceptance. It is therefore useful to determine, for a specified class of
theories and local expressions, whether a strictly gauge-invariant local
representative can exist at all. Here we address this question for the
linearised metric spin-2 gauge complex in four dimensions, under explicit
locality, polynomiality, dimension, derivative-order and regularity assumptions.
The results do not constitute a necessary-and-sufficient criterion for
arbitrary gauge theories.

Electromagnetism provides a useful comparison. The Bessel-Hagen
method~\cite{BesselHagen1921} supplements the field variation induced by a rigid
spacetime symmetry with an internal gauge transformation. In four-dimensional
Maxwell theory an appropriate compensation gives the gauge-invariant tensor
$\tau^\mu{}_\nu=-F^{\mu\alpha}F_{\nu\alpha}+\tfrac14\delta^\mu_\nu
F^{\alpha\beta}F_{\alpha\beta}$ directly from Noether's theorem, and all the
conformal currents are coordinate-weighted contractions of this same
tensor~\cite{BesselHagen1921,HobsonBarkerLasenbyWS}. The field
strength~$F_{\mu\nu}\sim\partial A$ is gauge invariant and contains one
derivative of the potential, so a quadratic expression in~$F$ has the dimension
and derivative order required of the local energy-momentum tensor.

In two companion papers we applied this construction to gravitational theories.
For Fierz-Pauli theory~\cite{FierzPauli1939,HobsonBarkerLasenbyFP}, the action
contains first derivatives of~$h_{\mu\nu}$, whereas the first local
gauge-invariant field strength is the linearised curvature,
$R^{(1)}\sim\partial\partial h$. The compensated form variation consequently
involves the linearised connection, and the construction yields a
gauge-invariant equivalence class of currents modulo divergences of
antisymmetric superpotentials and terms proportional to the field equations.
This equivalence is understood in the sense of the characteristic cohomology
of Barnich, Brandt and Henneaux~\cite{BarnichBrandtHenneaux2000}; it does not
select a preferred gauge-invariant local
tensor~\cite{DeserHenneaux1995,MagnanoSokolowski2002}. In particular, the
Poincar\'e currents considered in that analysis obey
\begin{equation}
 J^\mu_{\zeta,\xi}=J^\mu_{\zeta,0}-2\,\mathcal{E}^{\mu\nu}\xi_\nu,
 \label{eq:fpshift}
\end{equation}
where~$\zeta$ is the rigid Poincar\'e generator, $\xi$ is the compensating
spin-2 gauge parameter, and~$\mathcal{E}^{\mu\nu}$ is the linearised Einstein
tensor.\footnote{As in Ref.~\cite{HobsonBarkerLasenbyFP},
\cref{eq:fpshift} holds off shell for the natural representative of the
inexact gauge boundary term. Another representative adds an antisymmetric
superpotential~$\partial_\nu Y^{[\mu\nu]}$ to the right-hand side, without
changing the characteristic-cohomology class. Such a term appears explicitly
in the gauge variation~\labelcref{eq:currentdescent}.}
We show below that no admissible local compensator, or more general admitted
change of representative, makes the nontrivial translation class strictly
gauge invariant.

The derivative-order distinction in Fierz-Pauli theory motivates considering
an action constructed directly from a gauge-invariant curvature. Linearised
Weyl-squared gravity provides such an example: its action depends only on
$C^{(1)}_{\mu\nu\rho\sigma}$, the irreducible spin-2 part of the curvature.
As shown in Ref.~\cite{HobsonBarkerLasenbyWS}, however, the electromagnetic
outcome is still not obtained. Within the specified dimension-four quadratic
candidate space, invariance under both linearised diffeomorphisms and Weyl
rescalings permits only terms quadratic in~$C^{(1)}$ without further
derivatives. Four-dimensional identities reduce these to pure traces, whose
conservation requires them to vanish.

These two obstructions occur in theories for which constant spacetime
translations are also reducibility parameters of the gauge symmetry. A
reducibility parameter is a gauge parameter~$\epsilon$ for which the gauge
variation~$\delta_\epsilon\phi=\mathcal{R}\epsilon$ vanishes. In electromagnetism,
$\delta A_\mu=\partial_\mu\lambda$ vanishes for constant~$\lambda$, giving the
internal reducibility parameter associated with electric charge; a spacetime
translation is not such a parameter. For linearised gravity,
$\delta h_{\mu\nu}=2\partial_{(\mu}\xi_{\nu)}$ vanishes when~$\xi$ is a Killing
vector of the flat background,
$\xi_\mu=a_\mu+\omega_{\mu\nu}x^\nu$. Thus its constant part~$a_\mu$ can be
regarded either as a gauge parameter, for which
$\delta_{\xi=a}h_{\mu\nu}=0$, or as the generator of the rigid translation
$\Delta_a h_{\mu\nu}=-a^\rho\partial_\rho h_{\mu\nu}$.
The distinction between these actions is essential: the rigid symmetry has an
energy-momentum current, while the reducibility parameter determines a
surface-charge class. Although both constructions use the same constant
vector, reducibility alone does not imply the absence of a strictly invariant
local stress tensor.

We establish the non-existence result independently of this reducibility
observation, by enumerating the allowed rank-two tensors. The translation current is nevertheless nontrivial, since
rigid translations act nontrivially on the gauge-invariant curvature on the
regular solution sector. We then use homogeneity of the actions in total
derivative number to extend the result from the dimension and order of the
initial candidate space to every local polynomial representative satisfying
the remaining assumptions. This argument includes improvements involving
derivatives of the field equations. A separate compensator calculation
illustrates the difference from Maxwell theory, but is not needed for the
extension. Finally, we distinguish the quadratic energy current from the
linear surface form associated with a constant Killing parameter; relating
them requires the second-order field equation and Noether matching of a
consistent nonlinear completion.

The paper is arranged as follows. \Cref{sec:setup} defines the current
classes, local representatives and regularity conditions used throughout.
\Cref{sec:candidate} establishes the candidate-space result, together
with nontriviality of the translation current and the independent reducibility
charge. \Cref{sec:charge} gives the homogeneous-projection argument,
the compensator calculation and the distinction between the two classes.
We compare the resulting possibilities in \cref{sec:classification}, and
consider the degenerate Gauss--Bonnet and pure flat-space $R^2$ limits in
\cref{sec:weak}. The interpretation and relation to other constructions
are discussed in \cref{sec:discussion}. The parity-even conservation
matrix is recorded in \cref{app:kernel}.

We use the metric signature $(+\,-\,-\,-)$ and the curvature conventions of the
companion Weyl-squared paper~\cite{HobsonBarkerLasenbyWS}:
${R^\rho}_{\sigma\mu\nu}=\partial_\mu{\Gamma^\rho}_{\sigma\nu}
-\partial_\nu{\Gamma^\rho}_{\sigma\mu}
+{\Gamma^\rho}_{\lambda\mu}{\Gamma^\lambda}_{\sigma\nu}
-{\Gamma^\rho}_{\lambda\nu}{\Gamma^\lambda}_{\sigma\mu}$ and
$R_{\mu\nu}={R^\rho}_{\mu\rho\nu}$; with these conventions the four-dimensional
identities~\labelcref{eq:lovelock} and~\labelcref{eq:weylid} below carry their standard
coefficients. Symmetrisation and antisymmetrisation both carry unit weight,
$A_{(\mu\nu)}=\half(A_{\mu\nu}+A_{\nu\mu})$ and
$A_{[\mu\nu]}=\half(A_{\mu\nu}-A_{\nu\mu})$. We write $\mathcal{E}^{\mu\nu}$ for the
Euler--Lagrange (field-equation) expression of the theory under study, and $\eom$
for equality modulo the field equations. All statements are at linearised order
about flat space, where the gauge algebra of each theory is abelian.

\section{Bessel-Hagen currents and two notions of success}
\label{sec:setup}

Consider a linear gauge theory with fields~$\phi^i$, a quadratic action,
Euler--Lagrange expressions~$\mathcal{E}_i$ and gauge transformations
$\delta_\epsilon\phi^i=\mathcal{R}^i{}_\alpha\epsilon^\alpha$, where
$\mathcal{R}$ is a constant-coefficient differential operator (distinct from
the Ricci scalar~$R$). Noether's second theorem gives
$\mathcal{R}^{\dagger}{}_\alpha{}^i\mathcal{E}_i\equiv0$, with~$\mathcal{R}^{\dagger}$ the formal adjoint. A rigid translation~$a^\mu$ acts by
$\Delta\phi^i=-a^\mu\partial_\mu\phi^i$, and its canonical current satisfies
$\partial_\mu j^\mu=-\Delta\phi^i\mathcal{E}_i\eom0$.

The Bessel-Hagen prescription supplements this variation by a gauge
transformation, with the aim of obtaining a gauge-covariant combined variation.
Whether the current admits a preferred local representative is a further
question. We distinguish the following outcomes:
\begin{itemize}
\item \emph{Strong success}: the translation class has a representative
$-T^{\mu\nu}a_\nu$, where~$T^{\mu\nu}$ is local, strictly gauge invariant,
symmetric, conserved on shell and nontrivial. Nontriviality here means that
it is neither identically zero nor equivalent to a pure superpotential or
field-equation term. The Maxwell tensor has these properties.
\item \emph{Weak success}: gauge invariance holds only for the current class,
or a construction supplies only a superpotential expression, or an invariant
object has a different rank or dimension from the tensor sought. These are
distinct possibilities, grouped here only because none supplies the strong
representative just defined.
\end{itemize}
The Fierz-Pauli identity~\labelcref{eq:fpshift} displays the compensator freedom
within a current class; the non-existence arguments below establish that
strong success fails for that theory and for the regular pure
curvature-squared theories. The linear surface form associated independently
with a Killing parameter must be distinguished from a representative of this
translation class.

We require two properties of the quadratic theory. It is \emph{variationally
nontrivial} if its Euler--Lagrange operator is not identically zero. It is
\emph{dynamically regular} about the chosen background if, after quotienting
by the stated gauge transformations, the principal symbol has uniform rank
at generic momenta, with no degeneracy beyond those transformations, and the
linearised equations admit nontrivial finite-momentum local solutions. Here
the principal symbol is the highest-derivative part of the Euler--Lagrange
operator with~$\partial_\mu$ replaced by~$k_\mu$. Explicitly, modulo the
four-dimensional Gauss--Bonnet and Pontryagin densities the quadratic action
is
\begin{equation}
 S^{(2)}=\int\dd^4x\,\big[a\,C^{(1)}_{\mu\nu\rho\sigma}C^{(1)\mu\nu\rho\sigma}
 +b\,(R^{(1)})^2\big].
 \label{eq:quadaction}
\end{equation}
The regular sector is $a\neq0$. For $b\neq0$ the principal symbol, after
quotienting by the linearised diffeomorphisms, has rank six at generic
momenta, one spin-0 and five spin-2 directions; for $b=0$, Weyl-squared
gravity, it has rank five, and the missing direction is the additional
Weyl symmetry, by which one also quotients. For $a=0$ the rank drops to
one, a degeneracy not accounted for by any gauge symmetry: this is the
pure $R^2$ limit of \cref{sec:weak}. Gauss--Bonnet fails the first
condition and pure flat-space $R^2$ the second; both are discussed in
\cref{sec:weak}.

We also assume the standard regularity of the stationary surface and
completeness of the Noether identities, so that a local function vanishing
on shell is a combination of the field equations and their derivatives.
This is the regularity assumption used in local
cohomology~\cite{BarnichBrandtHenneaux2000}. The Noether and reducibility
correspondences invoked below are understood under these assumptions; an
explicit antifield calculation is not needed for the no-go proof.

For curvature-squared gravity in the normalisation~$[h_{\mu\nu}]=0$, the
action density and energy-momentum tensor have mass dimension four, while
$[R^{(1)}]=2$. In Maxwell theory, $[A]=1$, $[F]=2$ and
$[\mathcal{E}]=[\partial F]=3$, and conservation follows from
$\partial_\mu\tau^{\mu\nu}=-F^{\nu\alpha}\partial_\mu F^\mu{}_\alpha
=-F^{\nu\alpha}\mathcal{E}_\alpha\eom0$.
The dimensions agree: $[F]+[\mathcal{E}]=5=[\partial_\mu\tau^{\mu\nu}]$.
For the homogeneous fourth-order gravitational theories, however,
$\mathcal{E}\sim\partial^2(\text{curvature})$, so a product of curvature and
field equation has dimension six, whereas~$[\partial_\mu T^{\mu\nu}]=5$.
Thus conservation cannot take the same product form.

A stronger statement follows for the candidates considered here. Strict
invariance and dimension four restrict a quadratic tensor to curvature
bilinears, whose divergences depend only on second and third derivatives of
the metric perturbation. Those derivatives can be specified arbitrarily at
a point by cubic polynomial solutions of the homogeneous fourth-order field
equations. The lemma below consequently turns on-shell conservation into
identical conservation for this candidate space, a restriction that has no
counterpart for the Maxwell tensor, whose conservation uses the field equations.

\section{Direct localizability no-go}
\label{sec:candidate}

We first determine the space of strictly gauge-invariant local rank-two
tensors allowed by the dimension and derivative order of a pure
curvature-squared action in four dimensions. The conservation condition
reduces this space to the zero tensor. We then establish separately that
the translation current defines a nonzero class and that a constant Killing
translation determines a linear surface-charge form.

\paragraph*{On-shell and identical conservation.}
For the candidate tensors specified in the following lemma, conservation on
solutions of the homogeneous fourth-order field equations implies identical
conservation. This allows the candidate space to be determined by an
algebraic calculation.

\emph{Lemma.} Let a four-dimensional curvature-squared metric theory have a linearised
Euler--Lagrange operator homogeneous of fourth order in derivatives, so that every cubic
polynomial perturbation solves $\mathcal{E}_{\mu\nu}=0$ exactly. Let $T^{\mu\nu}$ be local,
polynomial and Poincar\'e-covariant in $h_{\mu\nu}$ and finitely many of its derivatives,
with no explicit $x$-dependence and constant dimensionless coefficients, symmetric, of
mass dimension four, quadratic in $h_{\mu\nu}$, and strictly invariant under linearised
diffeomorphisms. If $\partial_\mu T^{\mu\nu}\eom0$ then $\partial_\mu T^{\mu\nu}\equiv0$.

\emph{Proof.} By the invariant-jet
theorem~\cite{BoulangerDamourGualtieriHenneaux2001}, a strictly diffeomorphism-invariant
local polynomial in $h_{\mu\nu}$ is built from the linearised curvature and its
derivatives; since each curvature factor carries one field and two derivatives, a
dimension-four invariant quadratic in $h_{\mu\nu}$ is bilinear in \emph{undifferentiated}
linearised curvatures. Its divergence therefore depends only on the second and third
derivatives of $h_{\mu\nu}$ at a point. Every assignment of those jets is realised by a
cubic polynomial perturbation, whose fourth and higher derivatives vanish identically and
which by hypothesis solves the field equations exactly. On-shell conservation thus makes
the divergence vanish for every jet on which it depends, and so
$\partial_\mu T^{\mu\nu}\equiv0$.\nopagebreak\ $\qquad\blacksquare$

\paragraph*{No nonzero admissible candidate.}
A gauge-invariant, dimension-four, symmetric rank-two tensor quadratic in the
fields must be bilinear in the linearised curvature. Restricting first to
parity-even tensors, a complete spanning set is
\begin{align}
&T_1=R_{\mu\alpha}R_\nu{}^\alpha,\quad
 T_2=R\,R_{\mu\nu},\quad
 T_3=R^{\alpha\beta}R_{\mu\alpha\nu\beta},\notag\\
&T_4=R_{\mu\alpha\beta\gamma}R_\nu{}^{\alpha\beta\gamma},\quad
 T_5=\eta_{\mu\nu}R_{\alpha\beta}R^{\alpha\beta},\notag\\
&T_6=\eta_{\mu\nu}R^2,\quad
 T_7=\eta_{\mu\nu}R_{\alpha\beta\gamma\delta}R^{\alpha\beta\gamma\delta},
 \label{eq:basis4d}
\end{align}
where all curvatures are linearised. To see that this set is complete, the two
free indices of a curvature bilinear must occur either on the curvature
factors, giving $T_1$--$T_4$, or on $\eta_{\mu\nu}$ multiplying a curvature
scalar, giving $T_5$--$T_7$. The pair, cyclic and Bianchi symmetries reduce
all such contractions to the seven expressions above. By the lemma, a
conserved combination $T^{\mu\nu}=\sum_i a_iT_i^{\mu\nu}$ must satisfy
$\partial_\mu T^{\mu\nu}\equiv0$. Equating the coefficients of the independent
rank-one curvature-derivative monomials gives a rational linear system for
the $a_i$, with a one-dimensional kernel.\footnote{The independent monomials
are schematically of the form $R\,\partial R$. Their coefficients give a
rank-six system on the seven $a_i$, whose kernel is~\cref{eq:lovelock}.
The calculation and its symbolic verification are given in
\cref{app:kernel}.} The corresponding combination is
\begin{equation}
 8T_1-4T_2+8T_3-4T_4-4T_5+T_6+T_7,
 \label{eq:lovelock}
\end{equation}
which is the Lanczos--Lovelock combination arising in the variation of the
four-dimensional Gauss--Bonnet term. It is identically conserved by the
contracted Bianchi identity and vanishes identically in four
dimensions~\cite{Lanczos1938,Lovelock1971}, including both its trace and
trace-free part. Thus no nonzero conserved parity-even tensor remains.

For the parity-odd tensors, define
\begin{equation}
 \mathcal P=R_{\alpha\beta\gamma\delta}{}^\ast\!R^{\alpha\beta\gamma\delta},\qquad
 P_{\mu\nu}=R^{\alpha\beta}{}^\ast\!C_{\mu\alpha\nu\beta},
 \label{eq:oddbasis}
\end{equation}
where the dual acts on the first antisymmetric index pair. The complete parity-odd
candidate space is $aP_{\mu\nu}+b\,\eta_{\mu\nu}\mathcal P$. Here $P_{\mu\nu}$ is
symmetric by the pair-exchange symmetry of the dual Weyl tensor, trace-free by Weyl
tracelessness, and generally nonzero. Completeness follows by decomposing the algebraic
curvature representation as $(2,0)\oplus(0,2)\oplus(1,1)\oplus(0,0)$, the two Weyl
chiralities, trace-free Ricci and scalar curvature: its symmetric square contains exactly
one parity-odd scalar, from the two Weyl squares, and exactly one parity-odd symmetric
trace-free rank-two component, from Weyl times trace-free Ricci. These are $\mathcal P$
and $P_{\mu\nu}$. Weyl times Weyl contributes no trace-free rank-two target, mixed
chiralities giving $(2,2)$ and equal chiralities zero spin in one factor. The same count
shows that the antisymmetric targets $(1,0)$ and $(0,1)$ do not occur at all, so admitting
nonsymmetric rank-two candidates would add nothing.

The identity
\begin{equation}
 R_{\mu\alpha\beta\gamma}{}^\ast\!R_\nu{}^{\alpha\beta\gamma}
 =\tfrac14\eta_{\mu\nu}\mathcal P
 \label{eq:weylidodd}
\end{equation}
reduces that particular contraction to a pure trace, but does not eliminate the mixed
Ricci--dual-Weyl tensor.

Both coefficients are ruled out by explicit local solutions. With
$\eta=\operatorname{diag}(1,-1,-1,-1)$ and $\epsilon_{0123}=+1$, the perturbations
\begin{align*}
 \mathrm A:&\quad h_{00}=x^2,\quad h_{03}=h_{30}=y^3/6,\\
 \mathrm B:&\quad h_{00}=x^2,\quad h_{03}=h_{30}=xy^2/2,
\end{align*}
with all other components zero, are cubic and therefore solve the homogeneous
fourth-order linear equations exactly. At the origin, with components ordered as
$(t,x,y,z)$, direct substitution in exact arithmetic gives
\begin{equation}
\begin{array}{c|cc}
 &\partial^\mu P_{\mu\nu}&\partial_\nu\mathcal P\\ \hline
 \mathrm A&(0,-\tfrac12,0,0)&(0,0,0,0)\\
 \mathrm B&(0,0,0,0)&(0,0,-4,0).
\end{array}
\label{eq:oddtests}
\end{equation}
Conservation on A requires $a=0$, whereas conservation on B requires $b=0$.
There is therefore no nonzero conserved parity-odd candidate either.

Nor can the divergences of the two parity sectors cancel. By the lemma,
conservation of their sum is a polynomial identity in the second and third
metric derivatives. A reflection of these derivatives, with the free index
also transformed, preserves the parity-even contribution and reverses the
parity-odd one. Adding and subtracting the original and reflected identities
therefore requires each contribution to vanish separately.

Two special cases are instructive. For Weyl-squared gravity the tensor must in
addition be invariant under the linearised Weyl rescaling
$\delta_\omega h_{\mu\nu}=2\eta_{\mu\nu}\omega$, hence be built from the Weyl tensor
alone. The parity-even candidates are then spanned by
$C_{\mu\alpha\beta\gamma}C_\nu{}^{\alpha\beta\gamma}$ and $\eta_{\mu\nu}C^2$. In the
odd sector the Weyl contraction is the pure trace already treated
in~\cref{eq:weylidodd}, while $P_{\mu\nu}$ is excluded outright, since it is not Weyl
invariant: $\delta_\omega P_{\mu\nu}=-2(\partial^\alpha\partial^\beta\omega)
{}^\ast\!C_{\mu\alpha\nu\beta}\neq0$. Finally the four-dimensional Weyl
identity~\cite{HobsonBarkerLasenbyWS}
\begin{equation}
 C_{\mu\alpha\beta\gamma}C_\nu{}^{\alpha\beta\gamma}
 =\tfrac14\eta_{\mu\nu}\,C_{\alpha\beta\gamma\delta}C^{\alpha\beta\gamma\delta}
 \label{eq:weylid}
\end{equation}
reduces the first candidate to a pure trace, as found in the companion
paper. On the vacuum Ricci-flat solutions relevant to general relativity,
only $T_4$ and $T_7$ survive, since every other basis tensor contains a Ricci
factor. Moreover, $R_{\mu\nu\rho\sigma}=C_{\mu\nu\rho\sigma}$ there, so the
same identity reduces the rank-two super-energy to a pure trace. On this
Einstein subset the conserved gauge-invariant super-energy is instead given
by the rank-four Bel--Robinson tensor~\cite{Bel1958,Bel1962,Senovilla2000};
no rank-two counterpart survives.

The enumeration thus excludes every nonzero local, polynomial,
Poincar\'e-covariant, symmetric rank-two tensor that has dimension four, is
quadratic in $h_{\mu\nu}$, and is both strictly gauge invariant and
conserved. This conclusion is independent of a particular ansatz for the
Noether current. Its extension to the full admitted current class requires
the argument in \cref{sec:charge}.

\paragraph*{Nontriviality of the translation current.}
The translation symmetry is not trivial. Let $K^{(1)}$ denote the gauge-invariant
curvature, namely the linearised Riemann tensor $R^{(1)}$ in the purely
diffeomorphism-invariant theories or the linearised Weyl tensor $C^{(1)}$ in
Weyl-squared gravity. On the regular finite-momentum solution sector a translation
acts on it nontrivially, $\delta_a K^{(1)}=-a^\lambda\partial_\lambda K^{(1)}$. Gauge
transformations leave $K^{(1)}$ invariant and an on-shell-trivial transformation
vanishes on the stationary surface, so the translation is neither gauge nor on-shell
trivial; by the converse Noether correspondence, under the standard regularity
assumptions~\cite{BarnichBrandtHenneaux2000}, its $(n-1)$-current class is nonzero.

\paragraph*{The Killing surface-charge form.}
For the flat-background linearised metric equations considered here, the diffeomorphism
Noether identity reads $\partial_\mu\mathcal{E}^{\mu\nu}\equiv0$, and, by the algebraic
Poincar\'e lemma, the field equation is
therefore a double divergence of a
superpotential with the symmetries of the Riemann tensor. Hence for a constant Killing
$a_\mu$ the combination $\mathcal{E}^{\mu\nu}a_\nu$ is a total divergence,
\begin{equation}
 2\mathcal{E}^{\mu\nu}a_\nu=\partial_\alpha k^{\mu\alpha}_a[\mathcal{E}],
 \label{eq:komarcharge}
\end{equation}
for the corresponding linearised surface $(n-2)$-form $k_a$, namely the Komar/ADM form in the
Einstein case or an Abbott--Deser--Tekin/Iyer--Wald-type form for a general
higher-curvature theory~\cite{AbbottDeser1982,IyerWald1994,DeserTekin2002,
DeserTekin2003}. For the linearised Einstein operator this is explicit: with
\begin{equation}
 \mathcal{E}^{\mu\nu}=G^{(1)\mu\nu}=-\partial_\alpha\partial_\beta H^{\mu\alpha\nu\beta},
 \label{eq:komar}
\end{equation}
where $H^{\mu\alpha\nu\beta}=-H^{\alpha\mu\nu\beta}=-H^{\mu\alpha\beta\nu}
=\tfrac12(\eta^{\mu\nu}\bar h^{\alpha\beta}
-\eta^{\mu\beta}\bar h^{\alpha\nu}-\eta^{\alpha\nu}\bar h^{\mu\beta}
+\eta^{\alpha\beta}\bar h^{\mu\nu})$ and $\bar h_{\mu\nu}=h_{\mu\nu}
-\tfrac12\eta_{\mu\nu}h$, one has $k^{\mu\alpha}_a=-2a_\nu\partial_\beta
H^{\mu\alpha\nu\beta}$. This form is linear in $h$ and measures the charge of interior
sources, whereas the energy current is quadratic and measures the energy of the free
graviton; the two are brought into relation only by the second-order field equation of a
nonlinear completion, in the specific energy/source matching exhibited in
\cref{sec:discussion}. The class $[k_a]$ is moreover nonzero: under the stated
regularity and completeness assumptions, the standard correspondence between nontrivial
reducibility parameters and surface-charge classes assigns a nonzero class to a nonzero
constant translation parameter, since that parameter does not vanish on
shell~\cite{BarnichBrandtHenneaux2000}. For Weyl-squared gravity the relevant
full-complex parameter is the pair $(a_\mu,0)$; no classification of the conformal
Killing pairs is needed for this statement.

\paragraph*{Candidate-space Localizability Theorem.}
For each regular four-dimensional theory considered here, namely
Fierz-Pauli and the dynamically regular pure curvature-squared metric
theories, a nonzero constant translation $a_\mu$ has the following
properties: (i) its Noether current defines a nonzero class $[j_a]$;
(ii) this class has no strictly gauge-invariant rank-two representative in
the admissible tensor space; and (iii) as a Killing reducibility parameter,
$a_\mu$ determines a distinct linear surface-charge
form~\labelcref{eq:komarcharge}. For the curvature-squared theories, (ii)
follows from the enumeration above. For Fierz-Pauli it follows directly
from the invariant-jet theorem~\cite{BoulangerDamourGualtieriHenneaux2001}:
a strictly invariant quadratic expression requires two curvature factors
and hence at least four derivatives, whereas the action has only two.
Expressions such as $h\,\partial^2h$ have the required derivative order
but are not strictly invariant.

The translation energy is therefore nontrivial without admitting a strict
rank-two density in this tensor space. The linear surface charge associated
with the same parameter is a separate object. Gauss--Bonnet and pure flat
$R^2$ are degenerate limits excluded from the regular hypotheses, as
discussed in \cref{sec:weak}. We next extend the exclusion of strict
representatives beyond the prescribed tensor space.

\section{The current class and the reducibility charge}
\label{sec:charge}

The candidate space in \cref{sec:candidate} was restricted by the
dimension and derivative order of the action. To extend the result to all
admitted representatives of the translation current, we use the homogeneity
of the action to project a proposed representative onto the field degree
and total derivative number of the canonical current. The projection
preserves both its class and strict gauge invariance, including when the
improvement contains differentiated field equations. The proof uses the
nontriviality and regularity established above, but requires no explicit
cohomological descent. We subsequently examine the direct compensator
equation and distinguish the linear reducibility charge from the
quadratic current.

A local current is defined only modulo the equivalence
\begin{equation}
 j^\mu\sim j^\mu+\partial_\nu U^{[\mu\nu]}+(\text{field-equation terms}),
 \label{eq:currentclass}
\end{equation}
the Barnich--Brandt--Henneaux current class. Under a gauge transformation
$\delta_\xi h_{\mu\nu}=2\partial_{(\mu}\xi_{\nu)}$ the translation current varies as
\begin{equation}
 \delta_\xi j^\mu_a=+2\mathcal{E}^{\mu\nu}\Lambda_{a\nu}
 +\partial_\nu U^{[\mu\nu]}_a,\qquad
 \Lambda_{a\nu}=a^\rho\partial_\rho\xi_\nu,
 \label{eq:currentdescent}
\end{equation}
reproducing the Fierz-Pauli identity~\labelcref{eq:fpshift} and the Weyl-squared class
variation of the companion papers. The variation is thus trivial in the
class~\labelcref{eq:currentclass}, and the residual question is whether it can be made to
vanish \emph{strictly}.

\medskip
\noindent\textbf{Current-class Corollary.}\ \emph{For the theories of the
Candidate-space Localizability Theorem, whose actions are homogeneous quadratic with
total derivative number $q=2$ for Fierz-Pauli and $q=4$ for the pure curvature-squared
theories, and under the hypotheses used above, namely
locality, finite differential order, polynomial dependence on $h_{\mu\nu}$ with constant
coefficients, no explicit $x$-dependence, Poincar\'e covariance, linearity in the rigid
translation $a_\mu$, and the regularity of \cref{sec:setup}, no local representative
of the translation Noether class is strictly gauge invariant. The restriction to the
admissible tensor space is therefore inessential: the obstruction is a property of the
class.}
\medskip

\noindent\emph{Proof.} The actions considered here are homogeneous in total derivative
number, $q=2$ for Fierz-Pauli and $q=4$ for the pure curvature-squared metric theories.
Grade local polynomials by field degree $N$ and total derivative number $D$, each jet
$\partial_Ih_{\alpha\beta}$ carrying $N=1$ and $D=|I|$, and the constant parameter $a_\mu$
carrying $(0,0)$. A canonical translation current then has bidegree $(2,q)$, while the
linearised field equations have bidegree $(1,q)$ and their prolongations $(1,q+|I|)$, so
the stationary differential ideal is homogeneous. Let $J^\mu_a$ be a strictly
gauge-invariant representative of the class, so that
$J^\mu_a=j^\mu_a+\partial_\nu U^{[\mu\nu]}_a+W^\mu_a$ with $W_a\eom0$, and let
$\Pi_{2,q}$ denote projection onto bidegree $(2,q)$. Total differentiation preserves $N$
and raises $D$ by one, so
\begin{equation}
\begin{split}
 \Pi_{2,q}J^\mu_a&=j^\mu_a
 +\partial_\nu\big(\Pi_{2,q-1}U^{[\mu\nu]}_a\big)+\Pi_{2,q}W^\mu_a,\\
 &\qquad\text{with}\quad \Pi_{2,q}W_a\eom0,
\end{split}
 \label{eq:projequiv}
\end{equation}
the last term still weakly vanishing because the ideal is homogeneous: writing a general
element as $\sum f^{A,I}([h])\,\partial_I\mathcal{E}_A$, its $(2,q)$ part retains only
$|I|=0$ with $f$ linear in undifferentiated $h_{\mu\nu}$, so improvements built from
\emph{differentiated} field equations are covered. Hence
$K^\mu_a\equiv\Pi_{2,q}J^\mu_a$ represents the same class. It is again strictly
invariant, since the diffeomorphism variation maps bidegree $(N,D)$ to $(N-1,D+1)$ once
derivatives of the gauge parameter are counted, so contributions of distinct bidegree
cannot cancel one another; the Weyl variation maps $(N,D)$ to $(N-1,D)$ and likewise
preserves the decomposition, though diffeomorphism invariance alone suffices here.

By the invariant-jet theorem a strictly invariant local polynomial in
$h_{\mu\nu}$ is built from the linearised curvature and its
derivatives~\cite{BoulangerDamourGualtieriHenneaux2001}, each curvature factor carrying
one field and two derivatives. For $q=2$ no such bilinear exists, so $K_a=0$. For $q=4$,
$K_a$ is bilinear in undifferentiated linearised curvature, so $\partial_\mu K^\mu_a$
depends only on the second and third derivatives of $h_{\mu\nu}$. Every assignment of
those jets at a point is realised by a cubic polynomial perturbation, whose fourth
derivatives vanish identically and which therefore solves the homogeneous fourth-order
field equations exactly; on-shell conservation of $K_a$ is consequently \emph{identical}
conservation. The Superpotential Lemma below then gives
$K^\mu_a=\partial_\nu V^{[\mu\nu]}_a$ with $V$ local and polynomial. In either case the
translation class would be trivial, contrary to its nontriviality.\nopagebreak\ $\qquad\blacksquare$

\medskip
\noindent\textbf{Superpotential Lemma.}\ \emph{Let $K^\mu$ be a local, polynomial,
translation-invariant expression, quadratic in the fields, with
$\partial_\mu K^\mu\equiv0$. Then $K^\mu=\partial_\nu V^{[\mu\nu]}$ for a local
polynomial $V^{[\mu\nu]}$ with constant coefficients, no explicit $x$-dependence and, in
each homogeneous component, one lower total derivative number, though not necessarily a
lower highest jet order.}
\medskip

\emph{Proof.} Polarise $K$ into a symmetric bilinear differential operator on two
independent fields and replace the derivatives acting on the two factors by commuting
formal momenta $p$ and $k$, giving a polynomial symbol $\widehat K^\mu(p,k)$. The
hypothesis reads $r_\mu\widehat K^\mu=0$ with $r=p+k$. Decompose $\widehat K^\mu$ into
parts $\widehat K^\mu_m$ homogeneous of degree $m$ in $r$, with $s=p-k$ treated as
independent; each obeys $r_\mu\widehat K^\mu_m=0$. Set
\begin{equation}
 \widehat V^{\mu\nu}_m=\frac{1}{m+1}
 \left(\frac{\partial\widehat K^\mu_m}{\partial r_\nu}
      -\frac{\partial\widehat K^\nu_m}{\partial r_\mu}\right).
 \label{eq:superpot}
\end{equation}
Euler's identity gives $r_\nu\,\partial\widehat K^\mu_m/\partial r_\nu=m\widehat K^\mu_m$,
while differentiating $r_\nu\widehat K^\nu_m=0$ with respect to $r_\mu$ gives
$r_\nu\,\partial\widehat K^\nu_m/\partial r_\mu=-\widehat K^\mu_m$; adding,
$r_\nu\widehat V^{\mu\nu}_m=\widehat K^\mu_m$. For $m=0$ the hypothesis already forces
$\widehat K^\mu_0=0$. Summing over $m$ and restoring derivatives for $p$ and $k$ gives the
statement. The construction divides only by positive integers, uses no inverse
differential operator, lowers the total derivative number by one, and preserves Lorentz
covariance, linearity in $a_\mu$, constancy of the coefficients and the exchange symmetry
of the two field factors, which may in any case be imposed by averaging.\nopagebreak
$\qquad\blacksquare$

\medskip
\noindent Derivative homogeneity is essential to this proof. It applies
separately to Fierz-Pauli and pure curvature-squared metric theories, but
does not cover a mixed-order action such as Einstein-Hilbert plus
curvature-squared terms, for which the stationary ideal no longer has a
single derivative degree. We refer to the Candidate-space Localizability
Theorem and the Current-class Corollary together as the
\emph{Localizability Result}. Throughout, the admitted representatives
have finite differential order, are polynomial with constant coefficients
and no explicit coordinate dependence, and are Poincar\'e covariant.
Expressions involving inverse differential operators, nonanalytic or
pseudolocal dependence, or noncovariant improvements are outside its scope.

\paragraph*{The compensator mechanism.}
The failure of a direct compensator cancellation can be examined separately
and compared with the successful cancellation in Maxwell theory. Consider
an improvement containing undifferentiated field equations,
\begin{equation}
 j^\mu_a\longrightarrow j^\mu_a
 +\mathcal{E}^{\alpha\beta}\,\bar{Q}^\mu{}_{\alpha\beta}[h,a]
 +\partial_\nu V^{[\mu\nu]}_a ,
 \label{eq:charform}
\end{equation}
where $\bar{Q}^\mu{}_{\alpha\beta}=\bar{Q}^\mu{}_{(\alpha\beta)}$ is a local
functional. This includes, but is not restricted to, the Bessel-Hagen
choice $\bar{Q}^\mu{}_{\alpha\beta}=-2\delta^\mu_{(\alpha}\Psi_{a\beta)}$.
For example, $a^\mu h_{\alpha\beta}\mathcal{E}^{\alpha\beta}$ is an
improvement of the stated form that cannot be obtained from this choice.
The linearised field equations are symmetric and gauge invariant and,
under the regularity assumptions, furnish independent jet coordinates.
Consequently the explicit field-equation term in~\cref{eq:currentdescent}
is cancelled if
\begin{equation}
 \delta_\xi\bar{Q}^\mu{}_{\alpha\beta}
 =-2\,\delta^\mu_{(\alpha}\Lambda_{a\beta)} ,
 \label{eq:classshift}
\end{equation}
modulo $\eta_{\alpha\beta}(\cdots)$ in the Weyl case, where
$\mathcal{E}^\alpha{}_\alpha\equiv0$ makes any pure-trace part of $\bar{Q}$ invisible. On
this Bessel-Hagen choice the condition reads $\delta_\xi\Psi_{a\nu}=\Lambda_{a\nu}$.
The condition is sufficient for that cancellation but not necessary for strict invariance,
and the split of~\cref{eq:charform} into a characteristic term and an antisymmetric
divergence is not unique: a constant $\Lambda_{a\nu}$ makes
$\mathcal{E}^{\mu\nu}\Lambda_{a\nu}$ itself a superpotential divergence, either
by~\cref{eq:komarcharge} or by~\cref{eq:failedpeel} below applied with its parameter
replaced by $\Lambda_a$, whose derivative then vanishes. Nor do
improvements carrying \emph{derivatives} of the field equations reduce
to~\cref{eq:charform} by integration by parts alone: one integration leaves
$\partial_\lambda\big(Y^{\mu\lambda}{}_{\alpha\beta}\mathcal{E}^{\alpha\beta}\big)$,
whose tensor is not antisymmetric in $\mu\lambda$ and so is no admissible superpotential.
The following calculation concerns the sufficient cancellation
condition~\labelcref{eq:classshift}; the preceding corollary already treats the
full improvement freedom.

For the Bessel-Hagen choice of $\bar{Q}$, the transported parameter can be
separated into symmetric and antisymmetric derivatives,
\begin{equation}
 a^\rho\partial_\rho\xi_\nu
 =\tfrac12\,\delta_\xi\big(a^\rho h_{\rho\nu}\big)
 +a^\rho\partial_{[\rho}\xi_{\nu]},
 \label{eq:paramsplit}
\end{equation}
using $\partial_{(\mu}\xi_{\nu)}=\tfrac12\delta_\xi h_{\mu\nu}$. The
symmetric contribution is therefore supplied by the variation of
$\tfrac12 a^\rho h_{\rho\nu}$. The remaining term contains
$\partial_{[\rho}\xi_{\nu]}$, which in a symmetric-tetrad gauge is the
infinitesimal local Lorentz rotation accompanying the linearised
diffeomorphism. The metric perturbation has no antisymmetric component
whose variation could supply it. This interpretation suggests the
obstruction to direct cancellation; the lemma below establishes the
insolubility of~\cref{eq:classshift} for a general $\bar{Q}$ of any finite
derivative order.

To check the compensation explicitly, the most general Lorentz-covariant
zero-derivative expression is
\begin{equation}
 \Psi_{a\nu}=c_1\,a^\rho h_{\rho\nu}+c_2\,a_\nu h,
 \label{eq:psibasis}
\end{equation}
whose variation is
$\delta_\xi\Psi_{a\nu}=c_1a^\rho\partial_\rho\xi_\nu+c_1a^\rho\partial_\nu\xi_\rho
+2c_2a_\nu\partial_\rho\xi^\rho$. Matching the full target
$\Lambda_{a\nu}=a^\rho\partial_\rho\xi_\nu$ requires $c_1=1$ from the first structure and
$c_1=0$ from the absence of the second, which is inconsistent; equivalently, once the
symmetric half of~\cref{eq:paramsplit} has been absorbed, matching the remaining
$a^\rho\partial_{[\rho}\xi_{\nu]}=\tfrac12a^\rho\partial_\rho\xi_\nu
-\tfrac12a^\rho\partial_\nu\xi_\rho$ requires the same $c_1$ to be both $+\tfrac12$ and
$-\tfrac12$. The following lemma extends this inconsistency to a general $\bar{Q}$
of finite derivative order.

\medskip
\noindent\textbf{Compensator Lemma.}\ \emph{Let
$\bar{Q}^\mu{}_{\alpha\beta}=P^\mu{}_{\alpha\beta}{}^{\gamma\delta}(\partial,a)\,
h_{\gamma\delta}$ be any local functional with $P$ a constant-coefficient differential
operator of arbitrary finite order, of degree one in $a^\rho$, Lorentz covariant and
without explicit $x$-dependence. Then~\cref{eq:classshift} has no solution, with or
without the Weyl-case trace freedom, modulo additions to $\bar{Q}$ that are themselves
gauge invariant.}

\emph{Proof.} A $\bar{Q}$ of higher than linear order in $h$ would vary at nonzero order
in $h$, whereas the target of~\cref{eq:classshift} is $h$-independent; so $\bar{Q}$ is
linear in $h$, as assumed. On plane waves $h,\xi\sim e^{ik\cdot x}$ the variation is
$\delta_\xi\bar{Q}^\mu{}_{\alpha\beta}
=2P^\mu{}_{\alpha\beta}{}^{\gamma\delta}(ik,a)\,ik_\gamma\xi_\delta$, so the requirement
reads
\begin{equation}
 2P^\mu{}_{\alpha\beta}{}^{\gamma\delta}(ik,a)\,k_\gamma\xi_\delta
 =-2\,\delta^\mu_{(\alpha}(a\!\cdot\!k)\,\xi_{\beta)} .
 \label{eq:symbol}
\end{equation}
Decompose $P$ into parts homogeneous of degree $r$ in $k$, that is, with $r$ derivatives.
Each contributes at degree $r+1$ on the left, while the right-hand side is homogeneous of
degree one. Distinct degrees are independent as polynomials in $k$, so every $r>0$ part
must vanish or lie in the kernel of the gauge map, in which case it contributes nothing
and is a gauge-invariant addition. Only $r=0$ survives. The complete zero-derivative
basis, symmetric in $(\alpha\beta)$, linear in $h$ and of degree one in $a$, has seven
elements: $\delta^\mu_{(\alpha}a^\rho h_{\beta)\rho}$, $\delta^\mu_{(\alpha}a_{\beta)}h$,
$a^\mu h_{\alpha\beta}$, $\eta_{\alpha\beta}a^\rho h^\mu{}_\rho$,
$a_{(\alpha}h_{\beta)}{}^\mu$, $\eta_{\alpha\beta}a^\mu h$, and the parity-odd
$\epsilon^\mu{}_{\rho\sigma(\alpha}a^\rho h_{\beta)}{}^\sigma$; other $\epsilon$
contractions vanish, since $h_{\mu\nu}$ is symmetric and any $\epsilon$ carrying both
$\alpha$ and $\beta$ vanishes on symmetrisation. The resulting linear system for the seven
coefficients is inconsistent: its coefficient and augmented matrices have ranks $7$ and
$8$ respectively, so by Rouch\'e--Capelli there is no solution. Projecting
$(\alpha\beta)$ onto its trace-free part, which is the Weyl case, the two
$\eta_{\alpha\beta}$-carrying elements drop out and the ranks become $5$ and $6$: again
no solution. The Bessel-Hagen choice~\labelcref{eq:psibasis} uses the first two elements,
recovering the elementary contradiction above.\nopagebreak\ $\qquad\blacksquare$\footnote{The seven-term system,
its trace-projected variant with the ranks quoted, the degree separation on which the
lemma turns and the electromagnetic case below were all checked symbolically at
exact-rational jets.}
\medskip

\noindent The lemma rules out the direct cancellation
condition~\labelcref{eq:classshift} at every finite derivative order, and no
more: since that condition is not necessary for strict invariance, its
insolubility excludes no representative by itself. The exclusion of every
representative is the Current-class Corollary above. In Maxwell theory, by
comparison, $\delta_c A_\mu=\partial_\mu c$
and the transported scalar parameter $a^\rho\partial_\rho c$ is the
variation of $\lambda=a^\rho A_\rho$. Thus
$\delta_c\lambda=a^\rho\partial_\rho c$, the compensated form variation
is $a^\rho F_{\mu\rho}$, and the Maxwell tensor follows. There is no
antisymmetric pair of spacetime indices in this scalar parameter, unlike
the metric gauge parameter in~\cref{eq:paramsplit}. This explains the
difference between these two compensator constructions. It does not
apply directly to $p$-form gauge parameters, which are form-valued.

\paragraph*{The current and the reducibility charge.}
The constant Killing parameter $a_\mu$ independently determines a linear surface-charge
form $k_a$ of~\cref{eq:komarcharge}, the Komar/ADM charge, with $\dd k_a\eom0$; for
electromagnetism the same construction gives only the internal electric charge, not a
spacetime translation. It must not be identified with a superpotential for the
nontrivial translation current: an identity $j_a=\dd k_a+W_a$, with $W_a\eom0$, would
imply $[j_a]=0$, contradicting the nontriviality established above. Adding $\dd k_a$ as
a separate trivial improvement changes no current class. In the standard local
classification, $[j_a]\in H^{n-1}_{\rm char}(d)$ and $[k_a]\in H^{n-2}_{\rm char}(d)$,
corresponding respectively to antifield numbers one and two, or BRST ghost numbers $-1$
and $-2$~\cite{BarnichBrandtHenneaux2000}. Their quadratic and linear field degrees
provide an additional distinction within the homogeneous free theory. The relation
obtained after a nonlinear completion and its second-order field equation is discussed
in \cref{sec:discussion}.

The distinction can also be seen by attempting to express the
field-equation term in~\cref{eq:currentdescent} as a superpotential.
Writing $\mathcal{E}^{\mu\nu}=-\partial_\alpha\partial_\beta
H^{\mu\alpha\nu\beta}$ gives
\begin{equation}
 -\mathcal{E}^{\mu\nu}\xi_\nu
 =\partial_\alpha\big(\partial_\beta H^{\mu\alpha\nu\beta}\xi_\nu\big)
 -\partial_\beta H^{\mu\alpha\nu\beta}\,\partial_\alpha\xi_\nu ,
 \label{eq:failedpeel}
\end{equation}
so a remainder proportional to
$\partial_\beta H^{\mu\alpha\nu\beta}\partial_\alpha\xi_\nu$ remains.
For Fierz-Pauli, $H\sim h$ and this term has the form
$\partial h\,\partial\xi$, with a gauge-dependent first derivative of
the metric perturbation. For the curvature-squared theories,
$\mathcal{E}\sim\partial\partial R^{(1)}$ and $H\sim R^{(1)}$, so it
instead has the form $\partial R^{(1)}\partial\xi$, whose field-dependent
factor is gauge invariant. In either case it vanishes for a constant
parameter, for which $\partial_\alpha\xi_\nu=0$ and
$\delta_\xi h_{\mu\nu}=0$, but is generically nonzero otherwise. This
rearrangement therefore does not express the transported-parameter term
as the reducibility surface form.

There are two independent conclusions. Nontriviality of $[j_a]$ prevents
the energy current from being a local superpotential modulo the field
equations, whereas the Current-class Corollary excludes a strictly
invariant representative. The current and surface-charge classes may be
summarised as follows:
\begingroup\small
\begin{equation}
\begin{array}{c@{\hspace{1em}}c}
\underline{\text{codimension one}} & \underline{\text{codimension two}}\\[3pt]
[j_a]\ \text{translation current} & a_\mu\ \text{(Killing)}\\[2pt]
{\scriptstyle\rm var}\!\swarrow\ \ \searrow\!{\scriptstyle\rm cand} & \big\downarrow\\[2pt]
\underset{{\scriptstyle O(h\xi)}}{\delta_\xi j_a\neq0}\ \ \underset{{\scriptstyle O(h^2)}}{\{T^{\rm strict}_{\mu\nu}\}=\{0\}}
 & \underset{{\scriptstyle O(h)}}{k_a\ \text{surface charge}}
\end{array}
\label{eq:localization}
\end{equation}
\endgroup
The two columns refer to distinct classes of the free theory. An addition
$\dd k_a$ is a trivial current improvement and does not identify them.
Their physical relation requires the nonlinear completion and
energy/source matching discussed in \cref{sec:discussion}.

\section{Classification}
\label{sec:classification}

\Cref{tab:class} compares the theories considered here according to
whether translations are reducibility parameters and whether a strong local
stress representative exists.

\paragraph*{Internal theories.}
Maxwell theory, ordinary propagating free $p$-forms with
$1\leq p\leq n-2$, and Yang--Mills theory linearised about the trivial
vacuum all admit the standard strictly gauge-invariant stress tensor
quadratic in their field strengths. Their reducibility parameters are
internal gauge parameters; spacetime translations do not determine
reducibility charges. For Maxwell theory, the compensator calculation in
\cref{sec:charge} gives the invariant representative directly.
These examples do not establish that an internal gauge symmetry is
sufficient for strong success. Such a statement would require specified
field content, derivative order and invariant pairing, together with the
exclusion of topological and degenerate sectors such as Chern--Simons,
$BF$ and Gauss--Bonnet theories. We do not attempt that classification here.

\paragraph*{Construction independence.}
Since the Localizability Result concerns the current class, its conclusion
does not depend on the Bessel-Hagen prescription. As in the Fierz-Pauli
analysis~\cite{HobsonBarkerLasenbyFP}, this prescription is useful for
identifying the compensator obstruction, although it need not generate
every improvement. Canonical, Belinfante-improved and Hilbert expressions
belong to the same class when their differences are antisymmetric
superpotential divergences and field-equation terms. Relocalisation by
such a superpotential preserves the class by definition. Background
covariantisation, a change of pseudotensor prescription or another form of
relocalisation requires a separate check, since it need not preserve the
class.

\paragraph*{Massive theories.}
Proca and massive Fierz-Pauli theory have no gauge symmetry of the type
responsible for the present obstruction and admit local Hilbert/Belinfante
tensors. These examples likewise establish no converse theorem.
Gauss--Bonnet and pure flat $R^2$, on the other hand, retain the gauge
symmetry but fail the regularity assumptions; we consider them next.

\begin{table}[t]
\caption{Whether a strong local representative of the translation Noether class
exists, for the four-dimensional theories considered. Reducibility is always of the
diffeomorphism symmetry: a constant translation is a reducibility parameter of the
linearised diffeomorphisms, and not of the Weyl transformation
$\delta_\omega h_{\mu\nu}=2\eta_{\mu\nu}\omega$, which no constant $\omega\neq0$ leaves
invariant. Weyl-squared gravity carries that additional Weyl gauge symmetry, which is
what restricts its candidates to bilinears in $C^{(1)}$, which reduce to pure traces.
Gauss--Bonnet and pure flat
$R^2$ retain diffeomorphism reducibility but stand apart from the regular class of the
Localizability Result, being variationally trivial and dynamically degenerate
respectively.}
\label{tab:class}
\begin{ruledtabular}
\footnotesize
\begin{tabular}{lcc}
theory & transl.\ reducibility? & outcome \\
\hline
Maxwell, $p$-form, YM & no (internal) & strict $T_{\mu\nu}$ \\
Proca, massive FP & no (no gauge) & Hilbert/Belinfante \\
Fierz-Pauli & yes (diffeo) & no (connection) \\
Weyl-squared & yes (diffeo) & no (trace) \\
Gauss--Bonnet & yes, but var.\ trivial & degenerate \\
$R^2$ (flat) & yes, but dyn.\ irregular & degenerate \\
\end{tabular}
\end{ruledtabular}
\end{table}

\section{Degenerate limits: Gauss--Bonnet and pure $R^2$}
\label{sec:weak}

Gauss--Bonnet and pure flat-space $R^2$ gravity retain diffeomorphism
reducibility but fail different regularity assumptions of the
Localizability Result.

In four-dimensional Gauss--Bonnet gravity the linearised Euler--Lagrange
operator vanishes identically. This linear operator acting on
$h_{\mu\nu}$ should be distinguished from the quadratic curvature
expression~\labelcref{eq:lovelock}, which arises as the kernel of the
divergence map in \cref{app:kernel}. The action is topological
and its density is a total derivative, so the translation current has the
form $J^\mu=\partial_\nu k^{[\mu\nu]}$ and its charge is a boundary
integral. This does not require the local current density to vanish
pointwise or to be strictly invariant; variational triviality refers to
the identically vanishing Euler--Lagrange operator.

The distinction is relevant to the construction of Baker and
Kuzmin~\cite{BakerKuzmin2019}, who identify linearised Gauss--Bonnet
gravity as a completely gauge-invariant curvature-squared theory and
propose a correspondence with electrodynamics. Their energy-momentum
tensor is
\begin{equation}
\begin{split}
 T^{\mu\nu}_{\rm GB}={}&-R^{\mu\rho\lambda\sigma}R^\nu{}_{\rho\lambda\sigma}
 +2R_{\rho\sigma}R^{\mu\rho\nu\sigma}\\
 &+2R^{\mu\lambda}R^\nu{}_\lambda-RR^{\mu\nu}
 +\tfrac14\eta^{\mu\nu}\mathcal{G},
\end{split}
 \label{eq:bktensor}
\end{equation}
where $\mathcal{G}$ is the Gauss--Bonnet scalar formed from the linearised
curvature. A subsequent paper~\cite{BakerKuzmin2021} argues that the
action contributes to this tensor even though the four-dimensional field
equations vanish. In the spanning set~\labelcref{eq:basis4d}, however,
$T^{\mu\nu}_{\rm GB}$ is one quarter of~\cref{eq:lovelock}, or minus
one half of the conventional Lanczos tensor, and hence vanishes
identically in four dimensions. The proposed analogy with the Maxwell
tensor is therefore an algebraic correspondence between curvature and
field-strength bilinears, also noted in
Ref.~\cite{HobsonBarkerLasenbyWS}; it does not supply a nonzero
energy-momentum tensor.

For pure $R^2$ gravity about Minkowski space the degeneracy instead arises
from the dynamics. Its linearised field equation is
\begin{equation}
 \mathcal{E}_{\mu\nu}=2(\partial_\mu\partial_\nu-\eta_{\mu\nu}\Box)R^{(1)}
\end{equation}
which is gauge invariant and does not vanish identically. Taking the trace
gives $\Box R^{(1)}=0$, but substitution into the full equation imposes
the stronger condition $\partial_\mu\partial_\nu R^{(1)}=0$. Thus
$R^{(1)}$ is at most affine in the coordinates and vanishes under ordinary
localisation or finite-energy conditions. In particular, there is no
propagating finite-momentum scalar mode. The missing quadratic kinetic
term is conventionally interpreted as a strong-coupling limit; a scalaron
can reappear after adding an Einstein--Hilbert term, as in
$R+\alpha R^2$, or on a suitable curved background. One can nevertheless
form the tensor
\begin{equation}
 \Theta_{\mu\nu}=\partial_\mu\phi\,\partial_\nu\phi
 -\tfrac12\eta_{\mu\nu}\,\partial^\alpha\phi\,\partial_\alpha\phi,
 \qquad \phi\equiv R^{(1)},
\end{equation}
which obeys $\partial^\mu\Theta_{\mu\nu}=(\Box\phi)\partial_\nu\phi\eom0$
under the trace equation $\Box\phi\eom0$. Its dimension is six, rather
than the dimension four of the Noether tensor. More fundamentally, the
full equation $\partial_\mu\partial_\nu\phi=0$ removes the wave
solutions that would give this expression the interpretation of a scalar
stress tensor. Thus the weak success associated with a genuine scalar
energy-momentum tensor can occur in the nondegenerate completion
$R+\alpha R^2$, but not through a propagated scalaron in pure flat $R^2$.

\section{Discussion}
\label{sec:discussion}

The difficulty of defining local gravitational energy is familiar from
the non-tensorial character of gravitational pseudotensors and from the
surface-integral expressions for gravitational
charges~\cite{Komar1959,ADM1962,Wald1984,DeserTekin2002,DeserTekin2003}.
The present analysis establishes a precise local exclusion for the stated
linearised theories and distinguishes their translation current from
the reducibility surface charge. It thereby relates the Bessel-Hagen
construction to characteristic
cohomology~\cite{BarnichBrandtHenneaux2000} and to the known
non-uniqueness of linearised gravitational
currents~\cite{DeserHenneaux1995,MagnanoSokolowski2002,BakerKK2021,TaylorBaker2024}.

Taylor and Baker~\cite{TaylorBaker2024} provide a particularly relevant
comparison. They derive a general family of linearised gravitational
energy-momentum tensors by allowing boundary-equivalent Lagrangians and
translation-proportional field transformations motivated by the
Bessel-Hagen method. Some members are simultaneously symmetric and
traceless, although none is gauge invariant. Their classification leaves
open which field transformations are genuine symmetries of the action.
Here the Current-class Corollary excludes a strict representative under
all the admitted local polynomial improvements, while the compensator
lemma determines the failure of direct cancellation. Their examples also
show why symmetry and tracelessness, by themselves, do not establish
strict gauge invariance.

The compensator calculation has a useful interpretation in terms of the
antisymmetric parameter $\partial_{[\rho}\xi_{\nu]}$. Its contraction
with the rigid translation cannot be supplied by a variation of the
symmetric metric perturbation. In nonlinear gauge theories of gravity,
the corresponding local Lorentz factor acts covariantly on the fields,
and manifestly covariant Noether currents can accordingly be covariant
without being invariant. This is a suggestive relation between the two
constructions, rather than a derivation of the linearised obstruction
as a limit of the nonlinear theory. In particular, the mere presence of
a first-order field strength does not guarantee the Maxwell outcome.

The background and perturbative order also matter. Our result concerns
the local translation Noether current about Minkowski space. On a
constant-curvature background, for example anti--de Sitter space,
Abbott--Deser--Tekin/Iyer--Wald charges in quadratic-curvature theories
can be nonzero even when the corresponding asymptotically flat charges
vanish~\cite{AbbottDeser1982,IyerWald1994,DeserTekin2002,DeserTekin2003}.
These surface charges are compatible with the local exclusion considered
here.

A relation between the quadratic energy current and the linear surface
form arises after specifying a consistent nonlinear completion. Expanding
its diffeomorphism-invariant field equation as
$\mathcal{E}[g]=\lambda\mathcal{E}^{(1)}[h^{(1)}]+\lambda^2\big(\mathcal{E}^{(1)}
[h^{(2)}]+\mathcal{E}^{(2)}[h^{(1)},h^{(1)}]\big)+\cdots$ about
$g_{\mu\nu}=\eta_{\mu\nu}+\lambda h^{(1)}_{\mu\nu}+\lambda^2 h^{(2)}_{\mu\nu}+\cdots$
gives, at second order,
$\mathcal{E}^{(1)}[h^{(2)}]=-\mathcal{E}^{(2)}[h^{(1)},h^{(1)}]
\equiv t^{(2)}[h^{(1)}]$. The quadratic effective source constructed
from the first-order field therefore generates the second-order field.
Contraction with a background Killing vector and use of~\cref{eq:komarcharge}
give $a_\nu t^{(2)\mu\nu}=a_\nu\mathcal{E}^{(1)\mu\nu}[h^{(2)}]
=\tfrac12\partial_\alpha k^{\mu\alpha}_a[h^{(2)}]$. Thus the integral
of this source is the boundary charge obtained from the linear Komar/ADT
superpotential evaluated on the second-order field.

To identify $t^{(2)}$ with the free translation Noether class requires,
in addition, the usual Noether matching for the completion, up to
normalisation, improvements and terms proportional to the free field
equations. The second-order field equation alone establishes the relation
to the surface potential, but not this identification. General relativity
for Fierz-Pauli theory and nonlinear conformal gravity for Weyl-squared
theory provide the relevant completions. This Deser self-coupling
route~\cite{Deser1970} to ADM/ADT energy therefore uses information
beyond the free theory and its Localizability Result.

A recent proposal~\cite{Kenmoku2026} retains the second-derivative boundary
term of the Einstein--Hilbert density and applies the generalized canonical
formalism to claim a fully covariant gravitational energy--momentum tensor.
At linear order about flat space, however, the resulting leading contribution
is the identically conserved $O(h)$ superpotential
\begin{equation}
 t^{(1)\mu}{}_{\nu}=\frac{1}{2\kappa}\,
 \partial_\rho\big(\delta^\rho_\nu A^\mu-\delta^\mu_\nu A^\rho\big),
 \qquad
 A^\mu=\partial_\rho h^{\mu\rho}-\partial^\mu h,
 \label{eq:kenmokusuper}
\end{equation}
with $2\kappa$ the Einstein--Hilbert normalisation of~\cite{Kenmoku2026},
whose linearised-diffeomorphism variation,
\begin{equation}
 \delta_\xi t^{(1)\mu}{}_{\nu}=\frac{1}{2\kappa}\,
 \partial_\nu\big[\Box\xi^\mu-\partial^\mu(\partial\!\cdot\!\xi)\big],
 \label{eq:kenmokuvar}
\end{equation}
is generically nonzero. The construction therefore changes the current
representative by an $O(h)$ superpotential contribution, of the same
perturbative order as the surface-charge sector, but does not supply a
strictly gauge-invariant representative of the rigid-translation
stress-current class.

For a proposed energy-momentum tensor in a theory covered by the
Localizability Result, one can therefore ask whether it satisfies the
stated hypotheses and represents the translation class. If it does, it
cannot be strictly gauge invariant. Its superpotential contribution can
be examined at linear order, as above for Ref.~\cite{Kenmoku2026}.
Degenerate theories require their own analysis: in the Gauss--Bonnet
proposal~\cite{BakerKuzmin2019,BakerKuzmin2021}, the Lanczos identity
directly establishes that the tensor vanishes.

The curvature-squared examples also show that replacing an action
quadratic in first derivatives of the metric perturbation by one
quadratic in curvature does not restore the electromagnetic result.
Although the latter admits invariant tensors at the required derivative
order, their conservation excludes every nonzero candidate in four
dimensions. For Weyl-squared gravity this follows from reduction to pure
traces; for the general case it follows from the parity-even conservation
kernel and the two parity-odd tests. The degenerate Gauss--Bonnet and
pure $R^2$ limits do not evade this conclusion within its regular scope.
By comparison, Maxwell, propagating free $p$-forms and Yang--Mills about
the trivial vacuum possess first-derivative invariant field strengths
and strict stress tensors, while their spacetime translations are not
reducibility parameters. Three-dimensional Chern--Simons and
new-massive-gravity sectors would require a separate analysis, owing to
their dimensionful couplings, different curvature identities and possible
enhancement of the linearised symmetries.

Gauge theories of gravity with both internal and spacetime symmetries,
such as Weyl gauge theory and its extended
form~\cite{HobsonLasenbyBarker2024,CGTpaper,eWGTpaper}, raise a further
question. An effectively internal sector, for example the dilation
field-strength sector, might admit a strong local current of the
Maxwell type, whereas the diffeomorphism and local Lorentz sectors might
have local obstructions together with independent reducibility charges.
An independent connection, as in metric-affine or Poincar\'e gauge
gravity, supplies first-order gauge-covariant field strengths, so the
metric derivative-counting argument of \cref{sec:setup} no longer
applies in the same way.

One must also distinguish the rigid translations of the spacetime base
studied here from an internal $T(4)$ formulation, which defines a
gauge-source current soldered to the coframe. The latter is not, by its
construction alone, a representative of the external translation class.
Its invariance or covariance properties, and any relation to that class,
depend on the formulation. The corresponding construction in Weyl gauge
theory is developed in a companion paper~\cite{HobsonBarkerLasenbyT4}.

Finally, the candidate enumeration clarifies the role of the rank-four
Bel--Robinson tensor~\cite{Bel1958,Bel1962,Senovilla2000,PenroseRindler}.
The four-dimensional Weyl identities reduce the rank-two Weyl-square
tensors to pure traces, which exhaust the curvature bilinears on the
Ricci-flat subset. Their exclusion requires conservation as well as
these algebraic identities. Mixed Ricci--Weyl tensors are not generally
pure traces: the parity-even terms are treated by the conservation
kernel and the parity-odd $P_{\mu\nu}$ by~\cref{eq:oddtests}. On the
Einstein/Ricci-flat subset the Bel--Robinson tensor remains conserved,
so a gauge-invariant rank-four super-energy is available even though
there is no corresponding rank-two tensor. The Weyl identity responsible
for the rank-two reduction leaves the rank-four contraction intact.
This algebraic observation should be distinguished from conservation
and physical interpretation; we establish no analogous conserved
rank-four tensor for generic higher-curvature solutions.

\section{Conclusions}
\label{sec:conclusions}

We have established that the translation Noether current of Fierz-Pauli
and dynamically regular pure curvature-squared metric theories in four
dimensions has no strictly gauge-invariant local representative under
the stated assumptions. The candidate-space analysis gives the result
at the dimension and derivative order of the action: Fierz-Pauli has
too few derivatives to form a curvature bilinear, while in the
curvature-squared theories the parity-even conservation kernel yields
the identically vanishing Lanczos--Lovelock tensor and the parity-odd
coefficients both vanish. Homogeneity in total derivative number then
extends the exclusion to every finite-order local polynomial
representative with constant coefficients, no explicit coordinate
dependence and the stipulated covariance and translation dependence.
This includes weakly vanishing improvements containing differentiated
field equations.

The translation current nevertheless defines a nonzero class. The same
constant vector is independently a Killing reducibility parameter and
determines a linear surface-charge class. These classes have different
field degrees and occupy distinct positions in characteristic
cohomology, at antifield numbers one and two,
respectively~\cite{BarnichBrandtHenneaux2000}. Their relation as energy
and source requires a nonlinear completion, its second-order field
equation and the accompanying Noether matching. It does not follow from
the free theory alone.

The separate compensator calculation identifies why the Maxwell
cancellation cannot be repeated for the symmetric metric perturbation:
the transported gauge parameter contains an antisymmetric derivative
that cannot be supplied by its variation. This calculation illustrates
the difference between the constructions without being required for
the current-class proof. Maxwell theory, propagating free $p$-forms and
Yang--Mills about the trivial vacuum admit strict local stress tensors.
Gauss--Bonnet and pure flat $R^2$ instead lie outside the regular scope
of the result, through variational triviality and dynamical degeneracy,
respectively.

Independent-coframe and connection formulations raise a separate question
about local sources. In the companion internal-translation
construction~\cite{HobsonBarkerLasenbyT4}, the gravitational self-source
is local and strictly invariant under the introduced internal $T(4)$,
and its soldered density is invariant under the homogeneous internal
gauge factors in the scale-invariant formulation. These properties do
not by themselves identify that source with a representative of the
external rigid-translation class studied here. The distinction between
local currents, reducibility parameters and surface charges therefore
remains necessary when considering such nonlinear constructions.

\begin{acknowledgments}
W.~B. is grateful for the support of Girton College, Cambridge, Marie
Sk\l{}odowska-Curie Actions, and the hospitality of the Helsinki
Institute of Physics. Co-funded by the European Union (Physics for Future -- Grant Agreement No. 101081515). Views and opinions expressed are, however, those of the author(s) only and do not necessarily reflect those of the European Union or European Research Executive Agency. Neither the European Union nor the granting authority can be held responsible for them.
The authors used Claude (Anthropic; Fable 5.1) and ChatGPT (OpenAI;
GPT-6 Astra) for
checking analytical calculations, writing code for symbolic and exact-arithmetic
computations,
literature search, and drafting suggestions; all results were independently
verified, and the authors take full responsibility for the content.
\end{acknowledgments}

\appendix
\crefalias{section}{appendix}

\section{The four-dimensional conservation kernel}
\label{app:kernel}

For completeness, we give the linear system used to obtain the
parity-even conservation kernel~\labelcref{eq:lovelock}. Write
$T_{\mu\nu}=\sum_{i=1}^7 c_i T_{i\,\mu\nu}$ in the ordered spanning
set $T_1,\dots,T_7$ of~\cref{eq:basis4d}, with all curvatures
linearised. The contracted and second Bianchi identities express the
divergence in terms of six independent rank-one monomials,
\begin{align}
 M_1^\nu&=R_{\alpha\beta}\partial^\nu R^{\alpha\beta}, &
 M_2^\nu&=R\,\partial^\nu R,\notag\\
 M_3^\nu&=R^{\nu\alpha}\partial_\alpha R, &
 M_4^\nu&=R^{\alpha\beta}\partial_\alpha R_\beta{}^\nu,\notag\\
 M_5^\nu&=R_{\alpha\beta\gamma\delta}\partial^\nu R^{\alpha\beta\gamma\delta}, &
 M_6^\nu&=R^{\alpha\beta\gamma\nu}\partial_\alpha R_{\beta\gamma},
\end{align}
as $\partial^\mu T_\mu{}^\nu=\sum_{A=1}^{6}\big(\sum_{i=1}^{7}\mathsf{M}_{Ai}\,c_i\big)M_A^\nu$.
Identical conservation $\partial^\mu T_\mu{}^\nu\equiv0$ requires $\mathsf{M}c=0$; since
$\operatorname{rank}\mathsf{M}=6$, the $6\times7$ coefficient matrix (rows $M_1$--$M_6$,
columns $T_1$--$T_7$) and its one-dimensional kernel are
\begin{equation}
\begin{gathered}
 \mathsf{M}=\left(\begin{smallmatrix}
 0&0&1&0&2&0&0\\
 0&{\scriptstyle\frac12}&0&0&0&2&0\\
 {\scriptstyle\frac12}&1&0&0&0&0&0\\
 1&0&-1&0&0&0&0\\
 0&0&0&{\scriptstyle\frac12}&0&0&2\\
 0&0&-1&-2&0&0&0
 \end{smallmatrix}\right),\\[4pt]
 c\propto(8,-4,8,-4,-4,1,1),
\end{gathered}
 \label{eq:kernelmatrix}
\end{equation}
which is the Lanczos--Lovelock combination~\labelcref{eq:lovelock}. The seven
basis tensors obey the four-dimensional identity
$\sum_i c_i^{\rm Lov}T_i\equiv0$, so the one-dimensional coefficient
kernel represents the zero tensor. Combined with the exclusion of both
parity-odd structures~\labelcref{eq:oddbasis} by~\cref{eq:oddtests}, this
completes the candidate-space calculation. The matrix, its rank and its
null vector have also been reproduced by an independent exact symbolic
calculation.

\end{document}